\documentclass[useAMS,usenatbib]{mn2e}
\usepackage{times}
\usepackage[dvips]{epsfig}
\usepackage{color}
\usepackage{float}

\begin{document}
\label{firstpage}

\title[Evolution of Segue~1]{Could Segue~1 be a destroyed star cluster? --
  a dynamical perspective}

\author[Dom\'{i}nguez et al.]
{R. Dom\'{i}nguez$^{1}$\thanks{rdominguez@astro-udec.cl}, 
  M. Fellhauer$^{1}$, M. Bla\~{n}a$^{2}$, J.P. Farias$^{1}$,
  J. Dabringhausen$^{1}$, \newauthor G.N. Candlish$^{1}$,
  R. Smith$^{3}$ and N. Choque$^{1}$ \\ 
  $^{1}$ Departamento de Astronom\'{i}a, Universidad de Concepci\'{o}n,
  Casila 160-C, Concepci\'{o}n, Chile \\
  $^{2}$ Max-Planck-Institut für extraterrestrische Physik,
  Giessenbachstrasse 1, D-85748 Garching, Germany \\
  $^{3}$ Yonsei University, Graduate School of Earth System
  Sciences-Astronomy-Atmospheric Sciences, Yonsei-ro 50, Seoul
  120-749, Republic of Korea}

\pagerange{\pageref{firstpage}--\pageref{lastpage}} \pubyear{2016}

\maketitle

\begin{abstract}
  We attempt to find a progenitor for the ultra-faint object Segue~1
  under the assumption that it formed as a dark matter free star
  cluster in the past.  We look for orbits, using the elongation of
  Segue~1 on the sky as a tracer of its path.  Those orbits are
  followed backwards in time to find the starting points of our 
  N-body simulations.  The successful orbit, with which we can
  reproduce Segue~1 has a proper motion of $\mu_{\alpha} =
  -0.19$~mas\,yr$^{-1}$ and $\mu_{\delta} = -1.9$~mas\,yr$^{-1}$,
  placing Segue~1 near its apo-galacticon today.  Our best fitting
  model has an initial mass of $6224$~M$_{\odot}$ and an initial
  scale-length of $5.75$~pc.
\end{abstract}

\begin{keywords}
  galaxies: dwarfs --- galaxies: individual (Segue~1) --- methods:
  N-body simulations
\end{keywords}

\section{Introduction}
\label{sec:intro}

The Milky Way (MW) is surrounded by many dwarf spheroidal galaxies
(dSph).  With the advent of large surveys like for example the Sloan
Digital Sky Survey \citep[SDSS][]{yor00} many faint and also
ultra-faint dSph have been and are still discovered, increasing the
number of satellites of the MW tremendously \citep[see e.g.,][for one
of the latest discoveries]{lae15}.   

The high velocity dispersions observed in these objects suggest the
presence of a lot of dark matter (DM), if virial equilibrium and
standard gravity is assumed.  The assumption of virial
equilibrium enables to deduce a dynamical mass for these objects
\citep[e.g., chapter 4 in]{bin87}.  As these dynamical masses exceed
the visible mass in stars by orders of magnitude, the least luminous
dwarfs are considered the most DM dominated objects in the known
universe making them critical targets for indirect DM detection
experiments \citep[e.g.,][]{bal00,eva04,pie09,mar09}.   

Segue~1 was discovered by \citet{bel07} as an over-density of resolved 
stars in imaging data from the SDSS and the authors suggest that
Segue~1 is an extended globular cluster, possibly associated with the
Sagittarius stream.  This interpretation was contested by
\citet{geh09}, who demonstrated that the kinematics of stars in
Segue~1 clearly indicates that it is a dark matter dominated object.
They measured the radial velocities of $24$ stars in Segue~1 and
claimed that this object is a dwarf galaxy rather than a globular
cluster with a mass to light ratio of $1320$.  That makes Segue~1 one
of the most DM dominated objects known until this day.  \citet{nie09}
questioned this assumption.  If Segue~1 is a globular cluster that is
undergoing tidal disruption, then extra-tidal stars may not be so easy
to distinguish from gravitationally bound members.  Studies of
\citet[e.g.,]{smi13,bla15} have shown that objects on the brink of
destruction and/or close to apo-galacticon are surrounded by
sufficient extra-tidal stars to boost the measured velocity dispersion
by an order of magnitude or even more.  This boosted velocity
dispersion measurement, which does not represent the 'real' velocity
dispersion of a bound object in equilibrium, will lead to a
significant over-estimation of the dynamical mass and therefore to the
postulation of a heavily DM dominated object.  Furthermore, if
Segue~1 is immersed in the Sagittarius stream, the contamination of
any sample of Segue~1 stars by stars of the stream may be hard to
avoid.

Segue~1 is located at equatorial coordinates of $\alpha_{2000} \approx
151.77^{o}$, $\delta_{2000} \approx 16.08^{o}$ and at a distance of
$23$~kpc from the Sun \citep{bel07}.  Its central surface-brightness
is $\mu_{0} = 27.6^{+1.0}_{-0.7}$~mag\,arcsec$^{-2}$ and its
luminosity is about $340$~L$_{\odot}$ \citep{mar08}.  \citet{sim11}
after correcting for the influence of binary stars using repeated
velocity measurements, determined the velocity dispersion of Segue~1
to be $\sigma_{\rm los} = 3.7^{+1.4}_{-1.1}$~km\,s$^{-1}$ and the
radial velocity to be $v_{\rm rad} = 208.5 \pm 0.9$~km\,s$^{-1}$.  The
projected half-light radius is estimated to be $r_{\rm h} =
29^{+8.0}_{-5.0}$~pc.   

The goal of this project is to find a progenitor, which can reproduce
the observational data of Segue~1, mentioned above and shown in
Tab.~\ref{tab:obs}, under the assumption that Segue~1 has formed as a
star cluster, i.e., as an object without its own dark matter halo.
Success in this project does not mean that Segue~1 has to be a dark
matter free object, it simply opens up another possibility for
discussion.  In our simulations we do not include any information
about metalicities or star formation histories.  

In the next section we explain the setup of our simulations followed
by the description of our results.  We end this paper with some
conclusion and a brief discussion of our results. 

\begin{table}
  \centering
    \caption{Observables of Segue~1, we try to reproduce in this
    study. Rows 1 to 6 are taken from \citet{mar07}, rows 7 and 8 are
    taken from \citet{sim11}.}
  \label{tab:obs}
  \begin{tabular}{lcr}
    \hline
    Observable & Symbol & Value \\ \hline
    Right Ascension & $\alpha$ & $151.77^{o}$ \\
    Declination & $\delta$ & $16.08^{o}$ \\
    Distance & $D$ & $23 \pm 2$~kpc \\
    Central surface brightness & $\mu_{0}$ &
    $27.6^{+1.0}_{-0.7}$~mag\,arcsec$^{-2}$  \\
    Total luminosity & $L_{V}$ & $340$~L$_{\odot}$ \\
    Half-light radius & $r_{\rm h}$ & $29^{+8.0}_{-5.0}$~pc \\
    Velocity dispersion & $\sigma_{\rm los}$ &
    $3.7^{+1.4}_{-1.1}$~km\,s$^{-1}$ \\
    Radial velocity & $v_{\rm rad}$ & $208.5 \pm 0.9$~km\,s$^{-1}$ \\
    \hline 
  \end{tabular}
\end{table}

\section{Method \& Setup}
\label{sec:setup}

\begin{figure*}
  \centering
  \begin{tabular}{cc}
    \includegraphics[scale=0.5]{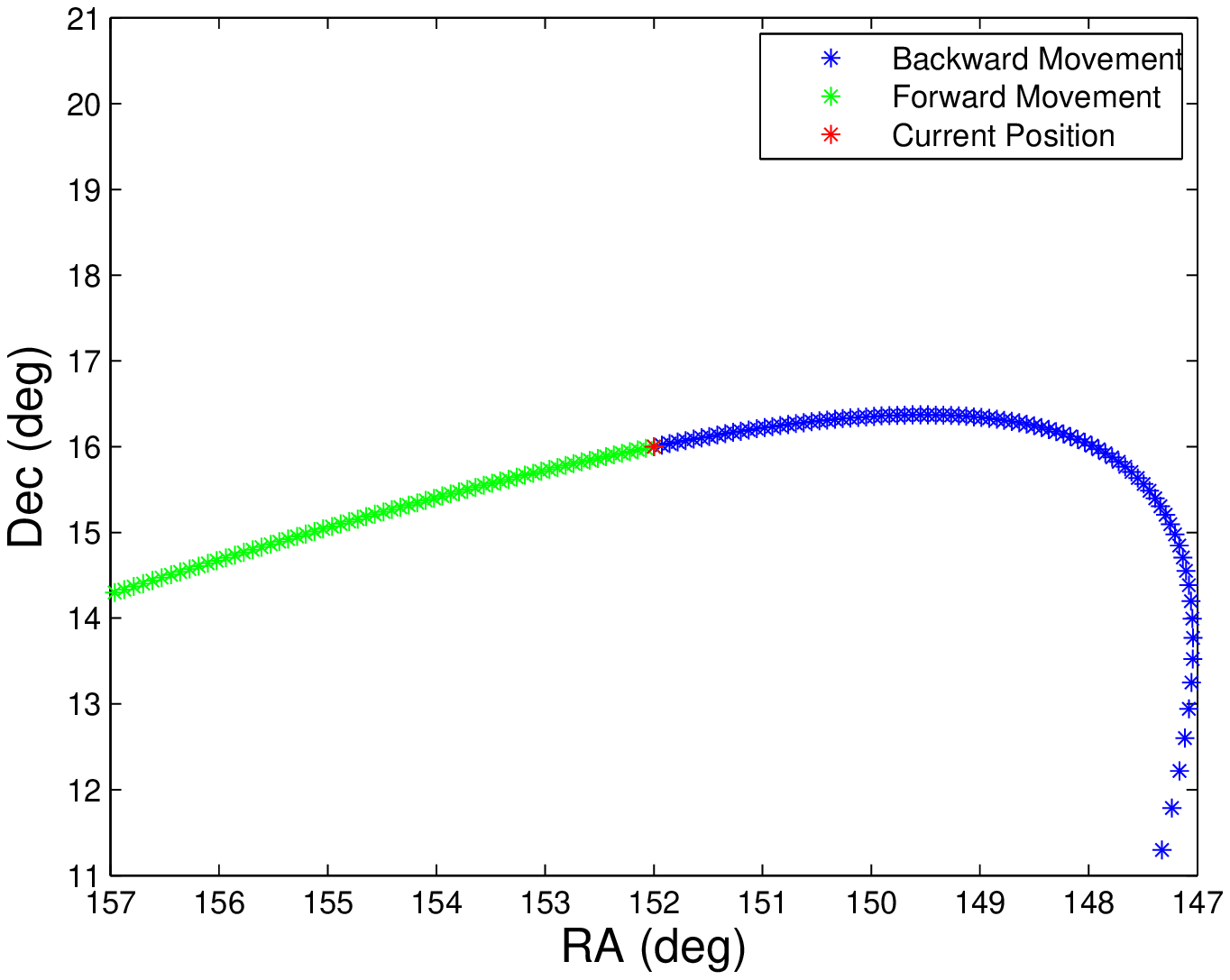}
    & \includegraphics[scale=0.50]{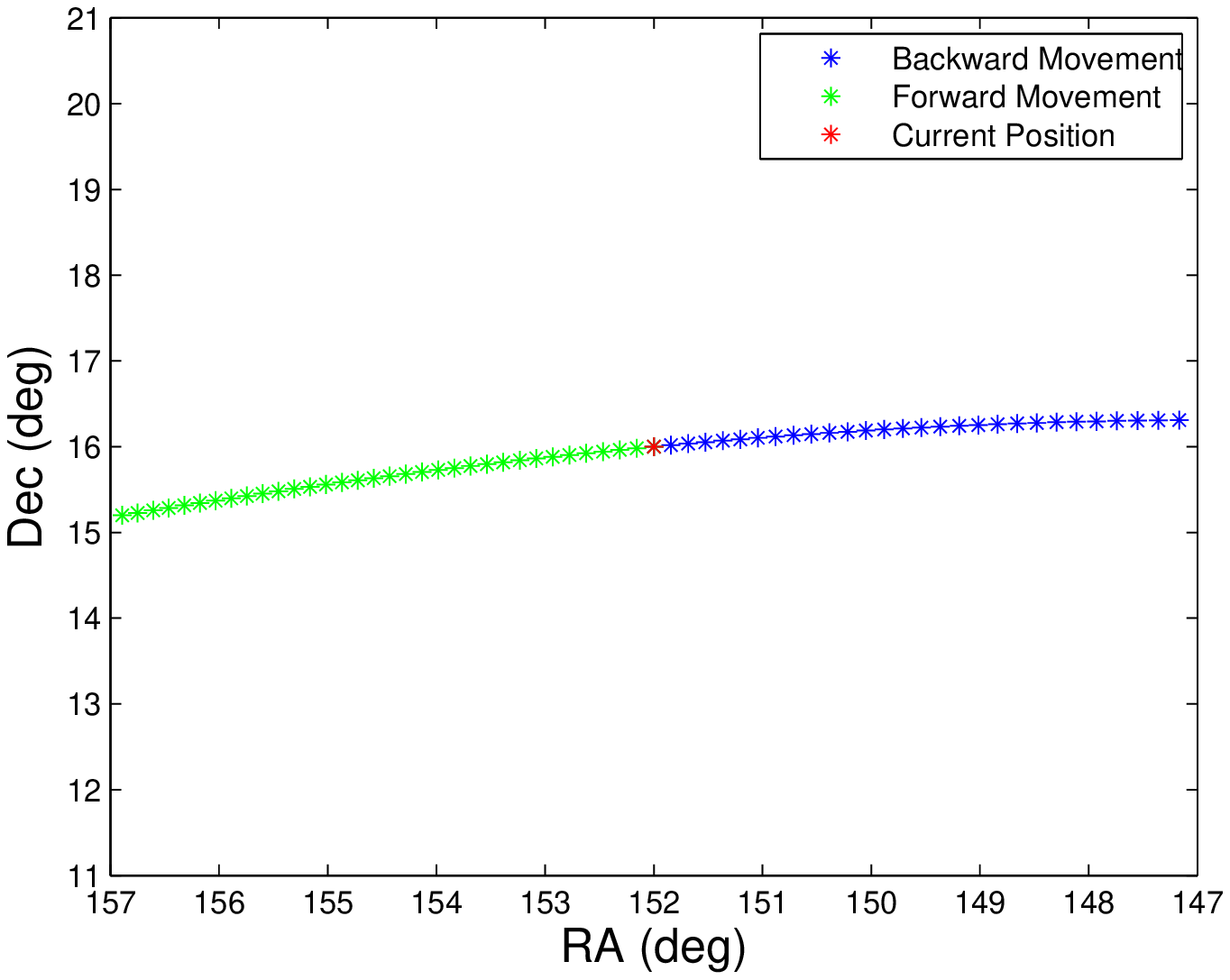} 
    \\\Large{(a)} & \Large{(b)}  
    \\\includegraphics[scale=0.50]{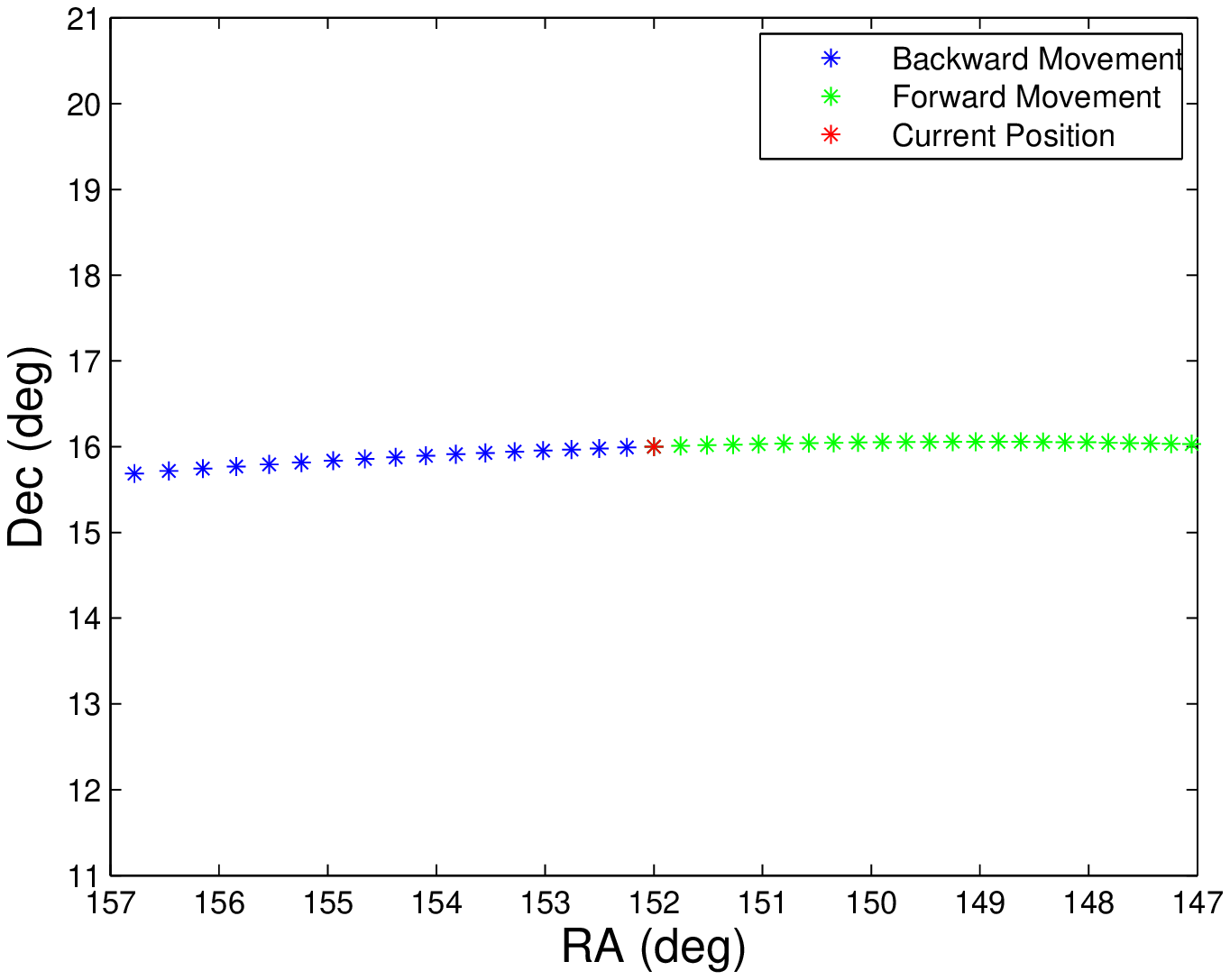}
    & \includegraphics[scale=0.50]{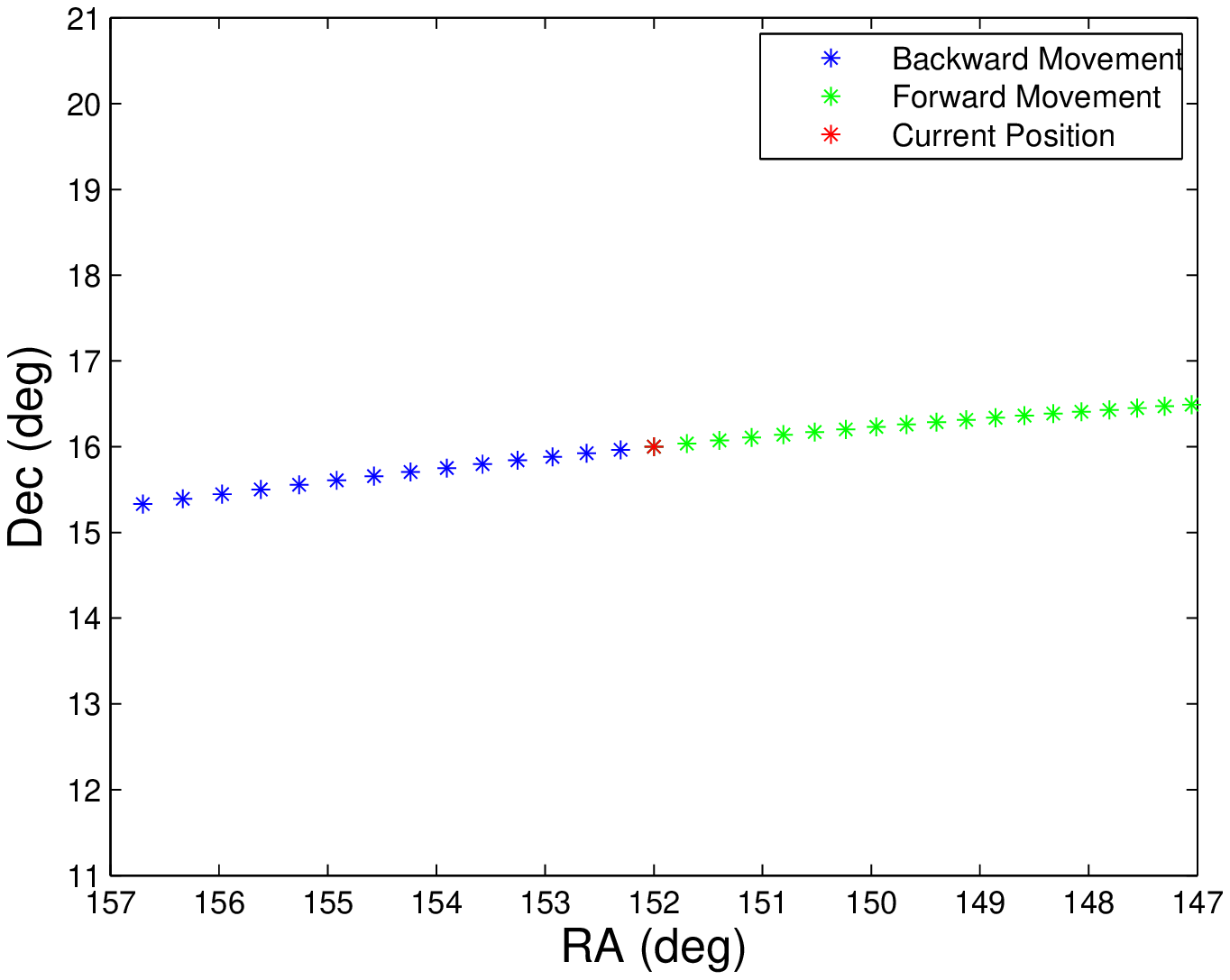} 
    \\\Large{(c)} & \Large{(d)}
  \end{tabular}
  \caption{Path of the possible orbits from Tab.~\ref{tab:pm} shown in 
    the sky in the vicinity of the actual position of Segue~1.  Blue
    stars show the orbit followed backwards in time, green stars show
    the future orbital path.  The red star denotes the position of
    Segue~1 today.  The time-step used in these test-particle
    calculations is 0.01~Myr.}   
  \label{fig:orbit}
\end{figure*}

\begin{table}
  \centering
  \caption{Some of the possible pairs of proper motions
    ($\mu_{\alpha}$ and $\mu_{\delta}$) matching the elongation of
    Segue~1.  First column indicates the panel of
    Fig.~\ref{fig:orbit}, where we plot the different orbits.  Second 
    and third column show the proper motions while fourth and fifth
    column give the peri-galacticon and apo-galacticon of the orbit,
    respectively.} 
  \label{tab:pm}
  \begin{tabular}{ccccc}
    \hline
    Orbit & $\mu_{\alpha}$ & $\mu_{\delta}$ & $R_{\rm peri}$ & $R_{\rm
      apo}$ \\
     & mas\,yr$^{-1}$ & mas\,yr$^{-1}$ & kpc & kpc \\ \hline
     a & -0.19 & -1.90 & 2.9 & 31.7 \\
     b & +0.10 & -1.90 & 5.6 & 31.4 \\
     c & -1.30 & -1.80 & 2.9 & 34.5 \\
     d & -1.50 & -1.70 & 5.0 & 35.8 \\ \hline
  \end{tabular}
\end{table}

\begin{table}
  \centering
  \caption{Position and velocity of Segue~1, today and at the start of
    the simulations, using orbit (a).} 
  \label{tab:pos}
  \begin{tabular}{crr} \hline
    Orbit (a) & Today & Start \\ 
    $t$ [Gyr] & 0 & -10 \\ \hline
    $x$ [kpc] & 19.15 & 11.95 \\
    $y$ [kpc] & 9.51 & -9.72 \\
    $z$ [kpc] & 17.73 & -0.40 \\
    $v_{x}$ [km\,s$^{-1}$] & 31.90 & 205.57 \\
    $v_{y}$ [km\,s$^{-1}$] & 49.79 & -112.90 \\
    $v_{z}$ [km\,s$^{-1}$] & 103.11 & -146.73 \\ \hline
  \end{tabular}
\end{table}

To determine a possible orbit for Segue~1, having only a measurement
of the radial velocity at hand, we assume pairs of proper motions
($\mu_{\alpha}$ and $\mu_{\delta}$) and perform test particle
integrations of the resulting orbit in a fixed analytic Milky Way
potential \citep[][and thereafter widely used as a standard
representation of the MW potential]{miz03}.  This potential is
parameterised as a logarithmic halo of the form:   
\begin{eqnarray}
  \Phi_{\rm{halo}}(r) & = & \frac{v_o^2}{2} ln(r^{2} + d^{2})
\end{eqnarray}
with $v_0 = 186$~km\,s$^{-1}$, $d = 12$~kpc and where $r$ is the
radius in kpc.

The disk is represented by a Miyamoto-Nagai potential: 
\begin{eqnarray}
  \Phi_{\rm{disk}}(R,z) & = & \frac{G M_{\rm{d}}} { \sqrt{R^2 + (b +
      \sqrt{z^2 + c^2})^2}}
\end{eqnarray}
with $M_{\rm{d}} = 10^{11}$~M$_\odot$, $b = 6.5$~kpc, $c=0.26$~kpc and
where $R$ is the radius within the plane in kpc and $z$ is the height
above or below the plane in kpc.  

Finally the bulge is modeled as a Hernquist potential: 
\begin{eqnarray}
  \Phi_{\rm{bulge}}(r) & = & \frac{G M_{\rm{b}}}{r+a}
\end{eqnarray}
using $M_{\rm{b}} = 3.4 \times 10^{10}$~M$_\odot$, $a = 0.7$~kpc and
where $r$ is the radius in kpc.  The superposition of these components
provides a reasonable representation of the Milky Way potential field
with a circular velocity at the solar radius of
$\sim~220$~km\,s$^{-1}$.  

\begin{figure*}
  \centering
  \begin{tabular}{cc}
    \includegraphics[width=8cm, height=8cm]{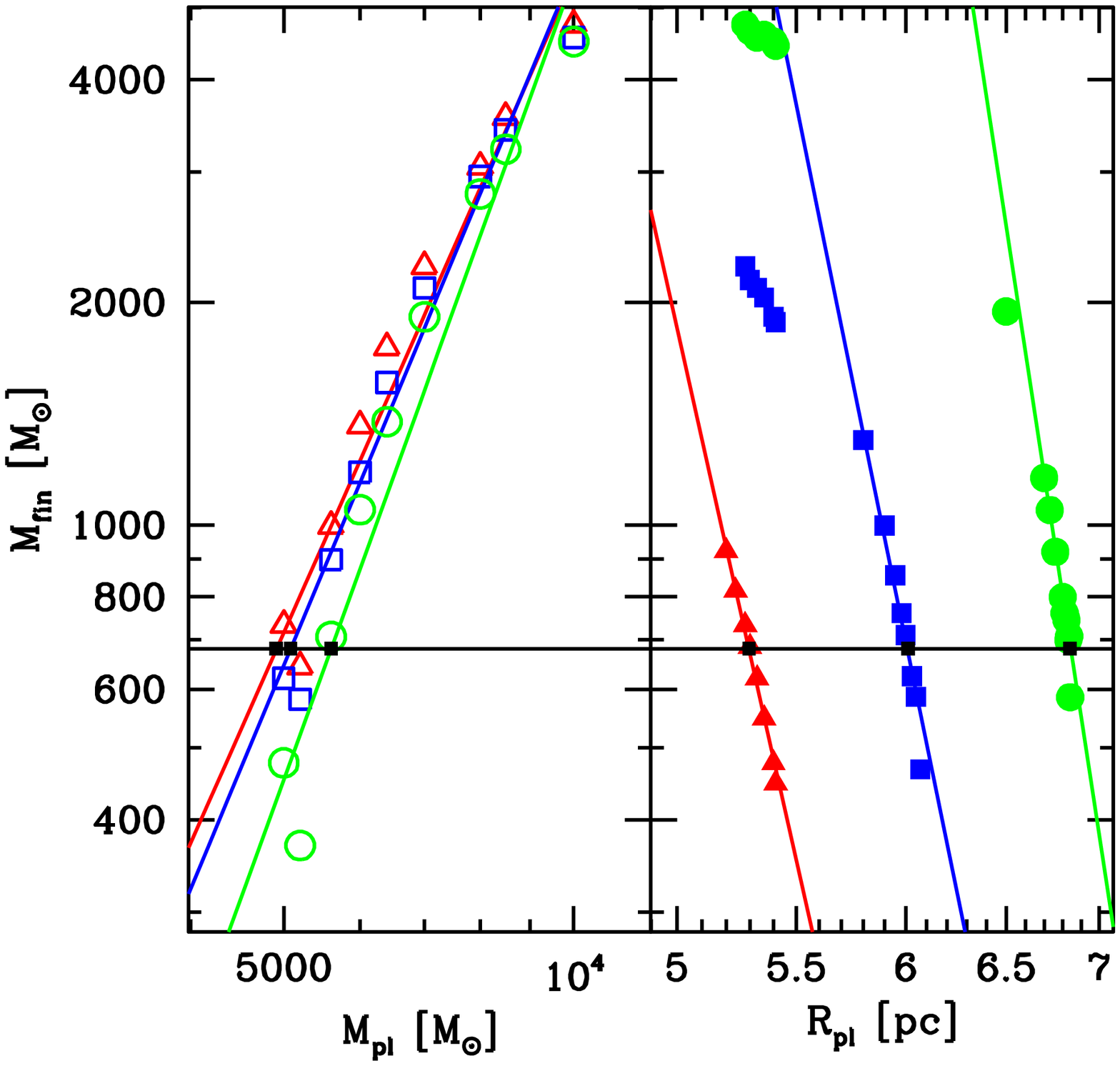}
    & \includegraphics[width=8cm, height=8cm]{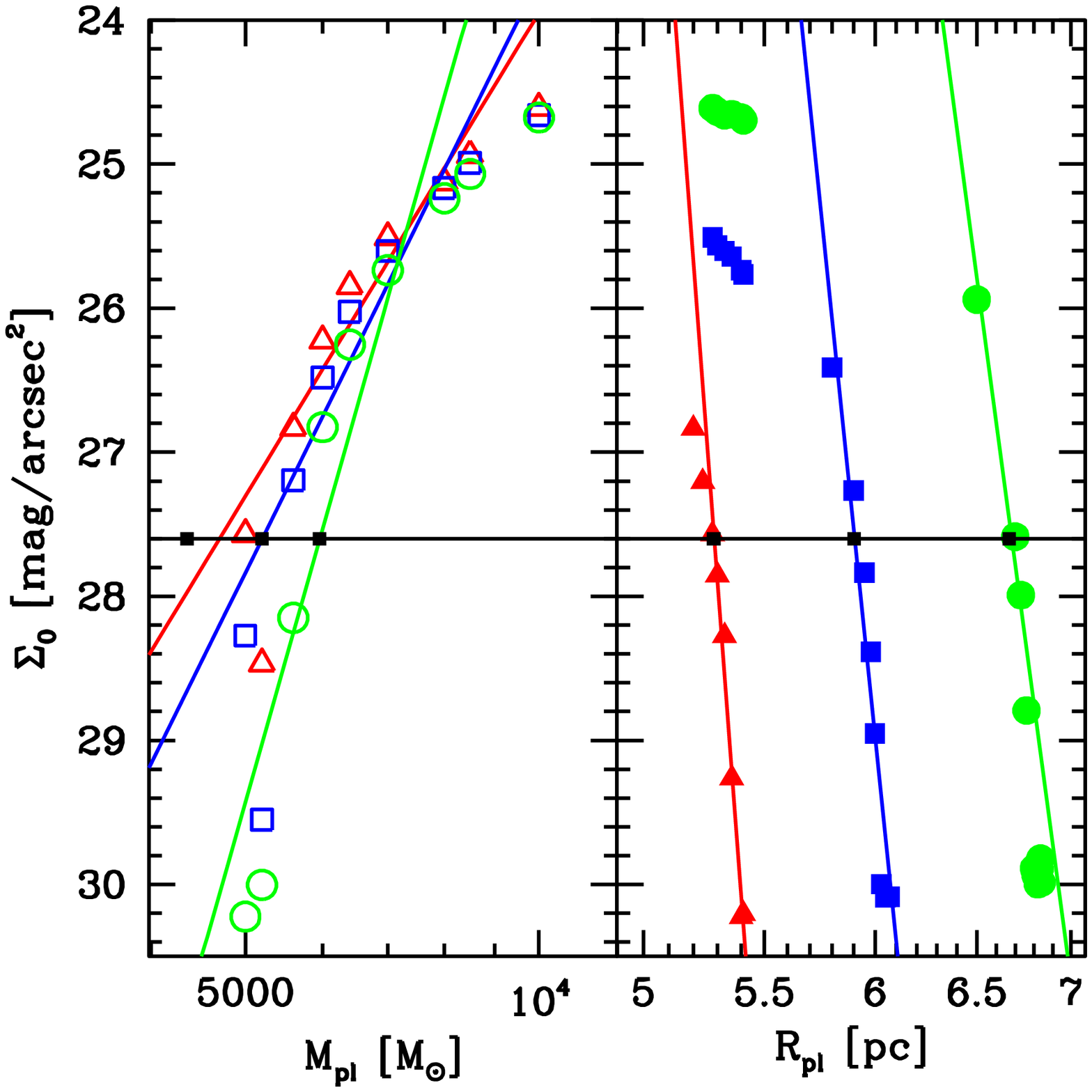} \\ 
    \includegraphics[width=8cm, height=8cm]{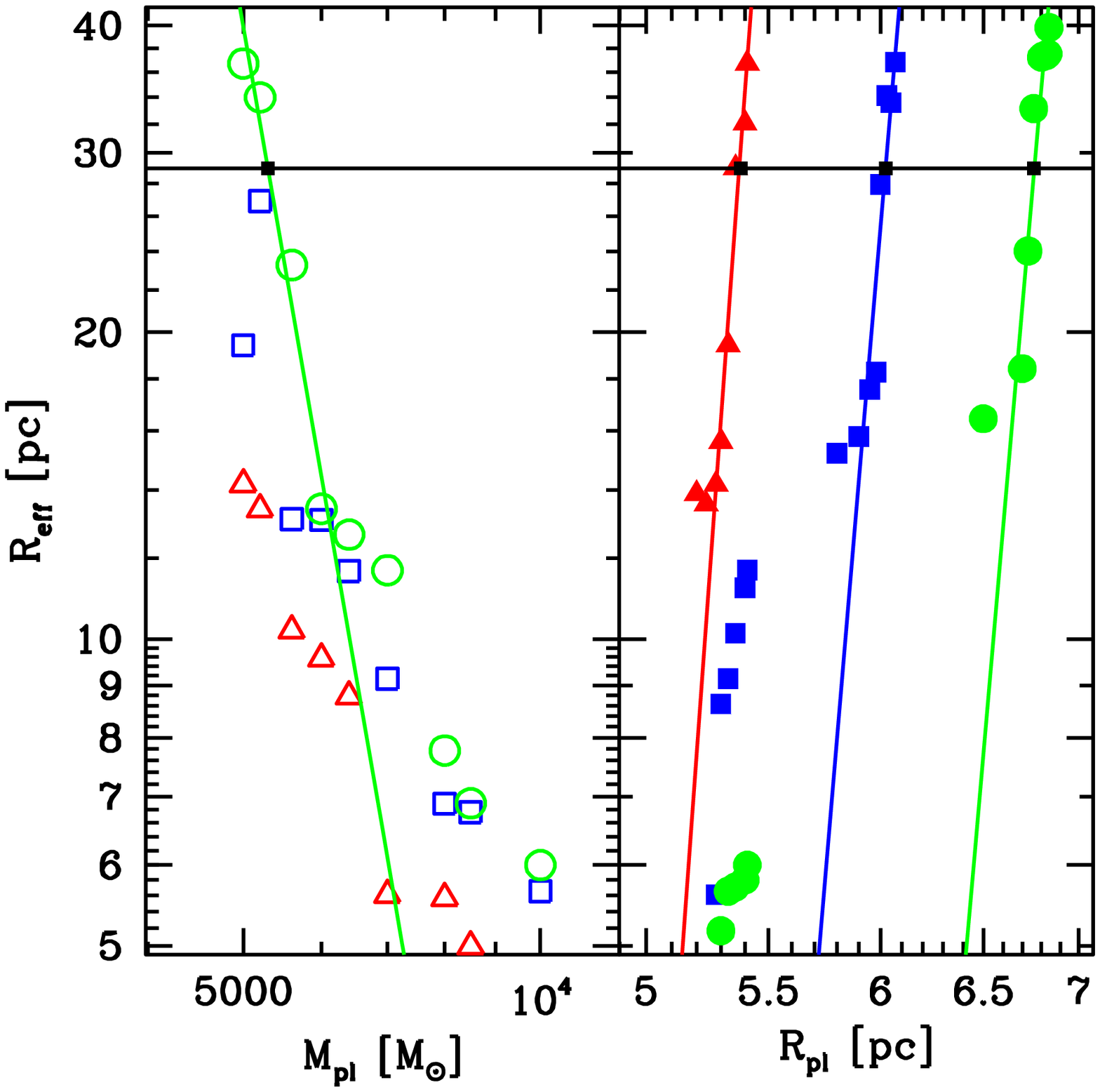}
    & \includegraphics[width=8cm, height=8cm]{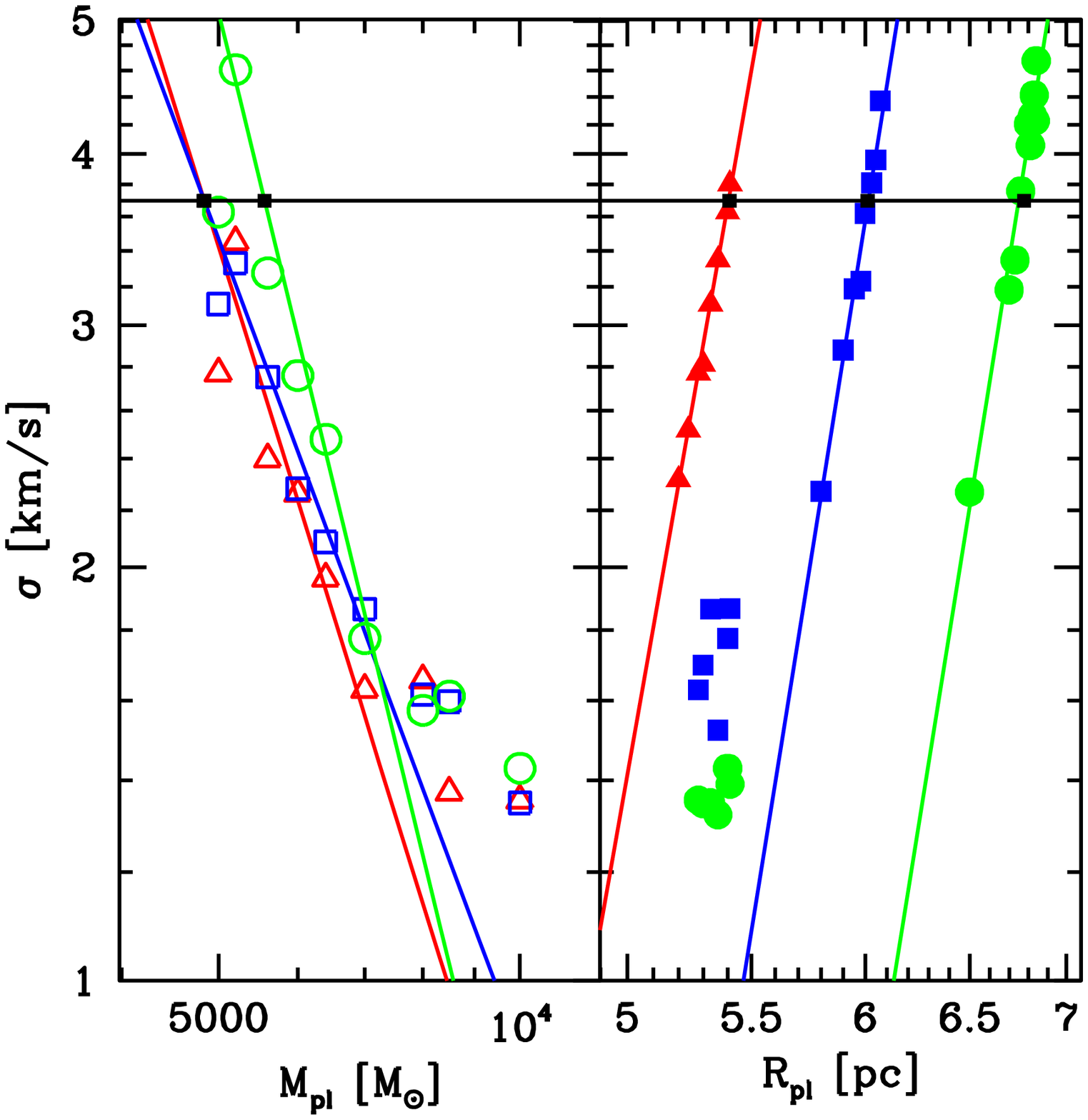}   
\end{tabular}
\caption{Final parameter (observable) as function of the initial
  Plummer mass $M_{\rm pl}$ for different choices of the initial
  Plummer radius $R_{\rm pl}$ (left part of each panel) and as a
  function of the initial Plummer radius for different choices of the
  initial mass (right part of each panel).  Top left: final mass, top
  right: central surface brightness, bottom left: effective radius,
  bottom right: velocity dispersion. Left side of each panel: we show
  the data points and the fitting lines for Plummer radii of 5.28 (red
  circles), 5.30 (blue stars), 5.33 (cyan squares), 5.40 (green,
  crosses), and 5.41 (magenta, diamonds) pc.  Right side of each
  panel: we show data points and fitting lines for initial masses of
  $1 \times 10^4$ (blue stars), $8 \times 10^3$ (red circles), $7
  \times 10^3  $ (cyan squares) and $5 \times 10^3$ (green
  crosses) solar masses .  The black lines indicate the upper and
  lower observational limit (Tab. 2) and the line in the center
  indicate the central observational value of Segue~1.  No
  observational error has been published for the mass so we only fit
  the inferred mass from the luminosity as was explained in section
  2.}
  \label{fig:mass}
\end{figure*}

At present, no proper motions are available for Segue~1.
\citet{nie09} tentatively presented further over-densities that may
belong to the object (their figure~4).  If we assume these patches are
in fact parts of the tidal tails of Segue~1, they map a path in the
sky, as the tails are assumed to align with the orbit.  This helps us
in restricting the possible pairs of proper motions.  We only consider
pairs of proper motions, which lead to orbits, whose path in the sky
is along these tentative over-densities.  Without the assumption,
i.e., that we know the path in the sky, any orbit would be possible. 
Furthermore, we restrict the possible orbits to solutions, which are
bound to the MW and discard first passages.

In Table~\ref{tab:pm} we present some of the possible pairs of proper
motions, which reproduce exactly the path in the sky outlined by the
patches of over densities in \citet{nie09}.   In Fig.~\ref{fig:orbit}
we show the projected orbits based on the proper motions from
Tab.~\ref{tab:pm} in the vicinity of the position of Segue~1 today.
Of course there are almost infinitely more solutions to this problem.
The orbits shown and used in this table have in common that Segue~1 is
close to its apogalacticon today.  This is the region of possible
orbits where we expect to find a solution.

We tried all of these orbits, using the method described below and
found that orbit (a) from Table~\ref{tab:pm} is the only one, where we
can find a possible initial star cluster evolving to a final object
which reproduces all observables given in Table~\ref{tab:obs}.   

Having established an orbit, we calculate this orbit backwards in time
for $10$~Gyr.  This backwards calculation is done using a
test-particle inside an analytic MW potential as described above.  

The choice for the length of this bachwards calculation in time is
rather arbitrary as we do not know the exact formation time
of Segue~1.  It represents a generic old object, orbiting the MW for a
long time.  Furthermore, it is clear that the potential of the MW was
not constant during the last $10$~Gyr, but rather growing with time
\citep[see for example the Via Lactea INCITE simulation of][]{kuh08}.
If we assume that the MW was less massive in the past and that Segue~1 
first orbited further out, closing in due to dynamical friction and
the growing potential of the MW, our generic choice of time together
with a constant MW potential represents an even longer time-span in
reality.  A more detailed description of this method and a discussion
about the simulation times used can be found in \citet{bla15} and
references therein.  

At the position (shown in Cartesian coordinates in
Table~\ref{tab:pos}) $10$~Gyr in the past we now insert a live model 
representing a possible stellar progenitor of Segue~1.  We model this
progenitor as a Plummer sphere \citep{plu11} with different Plummer
radii $R_{\rm pl}$ and initial masses $M_{\rm pl}$.  A Plummer sphere
is a widely used representation to model a young stellar cluster
\citep[e.g.,][]{boi03}.  We use the particle-mesh code {\sc Superbox}
\citep{fel00} to simulate this progenitor forward in time until we
reach again the position of Segue~1 today.  We analyse the final model
as described in Sec.~\ref{sec:res} and compare the data to the
observables.   

\begin{figure*}
  \centering
  \begin{tabular}{cc}
    \includegraphics[width=8cm, height=8cm]{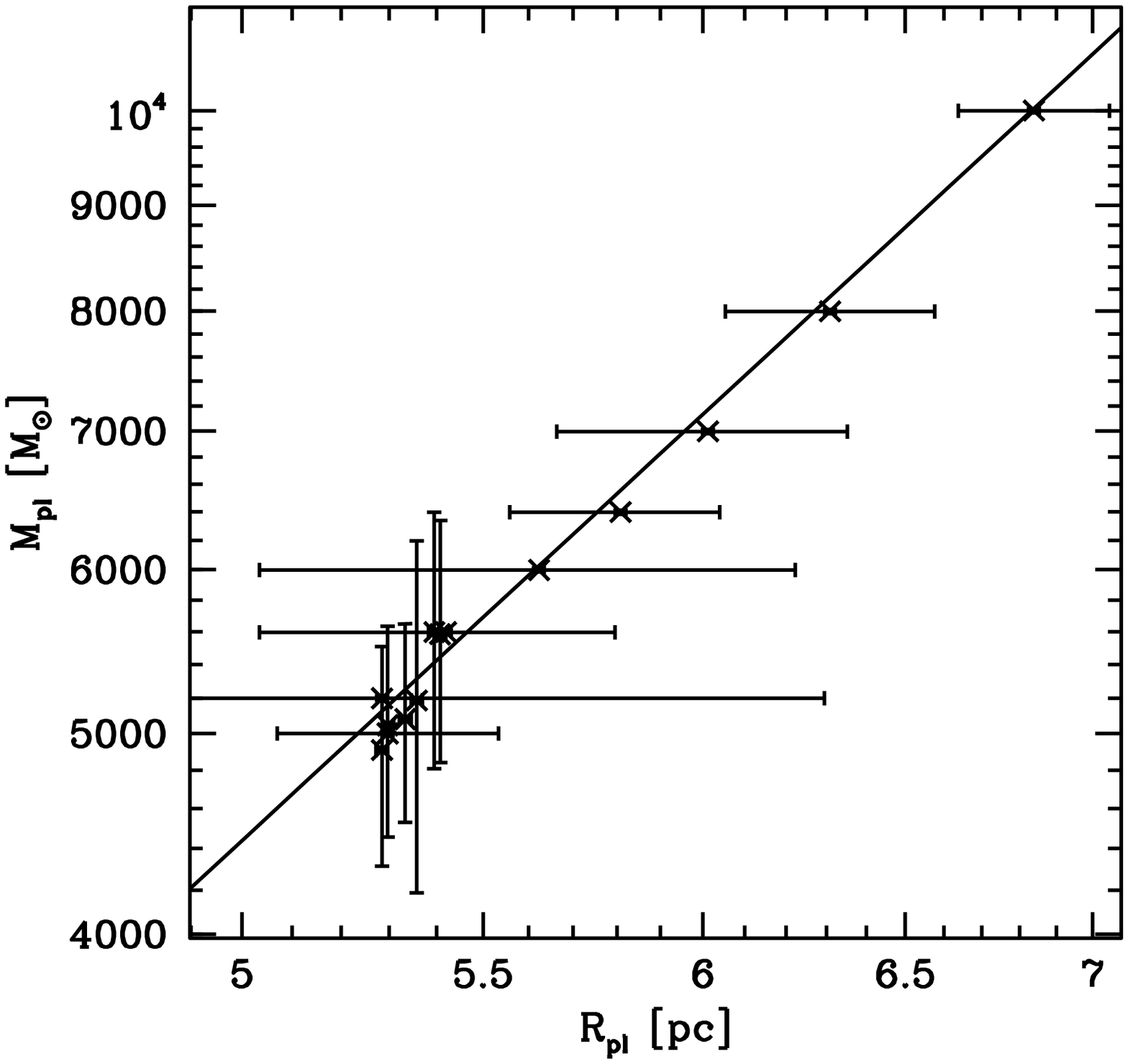}
    & \includegraphics[width=8cm, height=8cm]{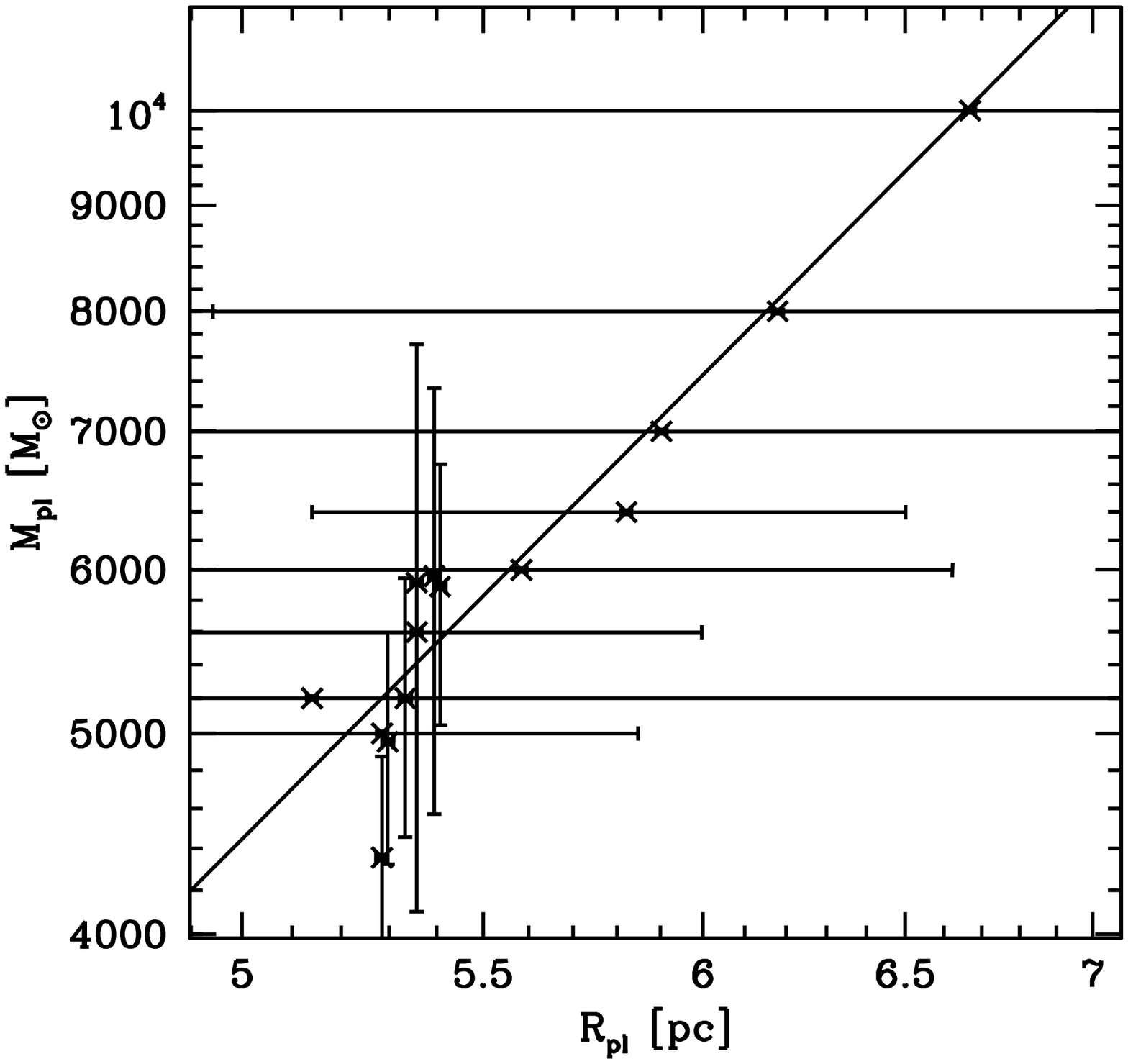} \\ 
    \includegraphics[width=8cm, height=8cm]{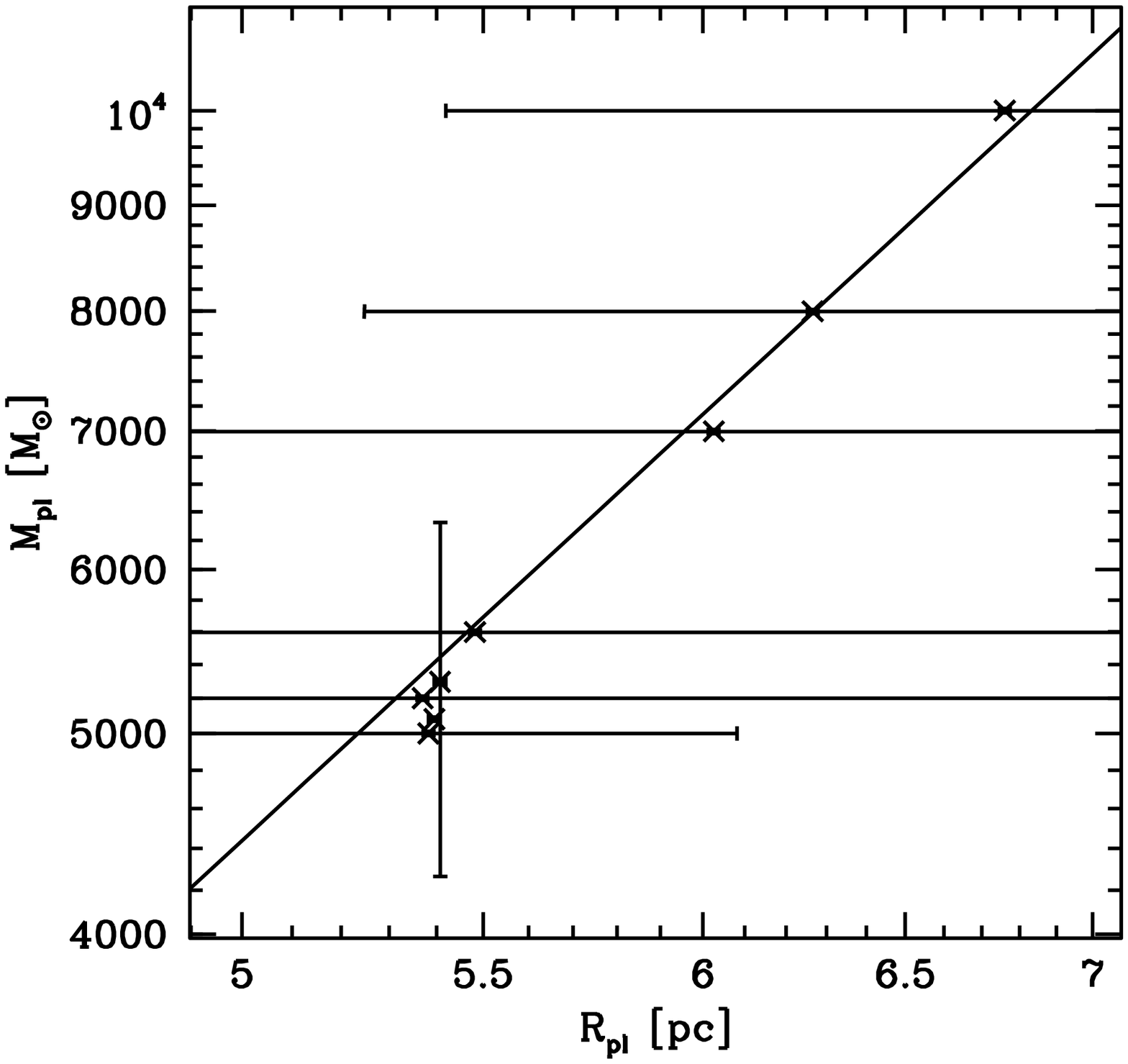}
    & \includegraphics[width=8cm, height=8cm]{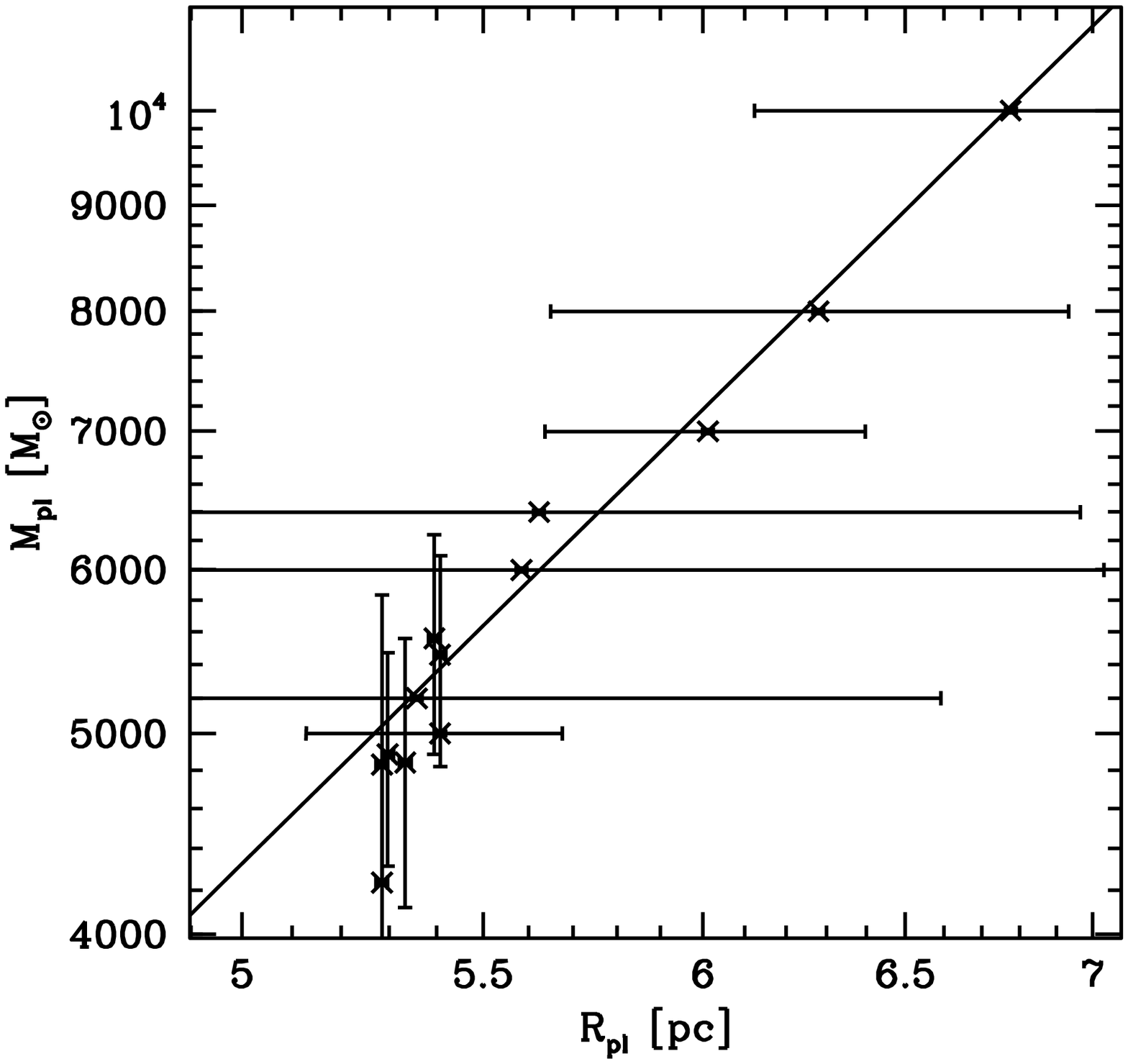} 
  \end{tabular} 
  \caption{Data-points show the pairs of initial parameters which lead
    to the correct final value of the given observable.  The
    data-points can be fitted by power-laws, describing 1D sub-sets
    of the initial 2D parameter space.  Top left: final mass, top
    right: central surface brightness, bottom left: effective radius,
    bottom right: velocity dispersion.} 
  \label{fig:power}
\end{figure*}

As we are simulating masses and not luminosities we convert our masses
into luminosities using a generic stellar mass-to-light ratio of $2 \
{\rm M}_{\odot}/{\rm L}_{\odot}$, e.g.,\ we assume the total stellar
mass of Segue~1 to be $680$~M$_{\odot}$.  This assumption is valid as
Segue~1 consists of old and metal-poor stars \citep{fre14}, which have
a mass-to-light ratio larger than unity.

To find a suitable progenitor model we follow closely the method
described in \citet{bla15}.  This method uses the fact that the final
mass of the object, its central surface brightness, its effective
radius and its internal velocity dispersion can each separately be
matched by a 1D subset, i.e., a power-law function, of the
initial 2D parameter space spanned by the parameters $M_{\rm pl}$ and
$R_{\rm pl}$.  Furthermore, it relies on the fact that each observable
has a different dependency on the initial parameters, i.e., the
power-laws are different for different observables.  If there is a
matching solution to the problem at hand, i.e., here to find a
possible progenitor for Segue~1, all these power-law functions should
ideally intersect in one point.  Due to measurement uncertainties, the
power-law functions will in practice not intersect in a single point,
but in the same small region in the parameter space of initial radii
and masses if there is a solution to our problem.  

In contrast to other methods searching for the correct initial
conditions, here, the search for a solution is performed over a wider
region of parameter space instead of using an educated first guess and
closing in on to the solution by trial and error.  This does not
necessarily mean that one has to perform less simulations in total.  

\section{Results}
\label{sec:res}

From the observables listed in Table~\ref{tab:obs}, the position,
i.e., the right ascension, declination and distance to the Sun, and
the radial velocity of the system are matched automatically by design
of the simulations.  The objects are simulated forward in time using a
particle-mesh code.  Obviously, the simulated objects should end up at
the same position today, where we started our backwards test-particle
calculation from.  Small deviations are to be expected as a live,
extended system behaves slightly different than a test-particle.
These deviations (in our simulations in all three dimensions) are by
an order of magnitude less than the observational uncertainty in the
measured distance. 

If we want to describe the results of our simulations we have to
distinguish between how a single simulation evolves in time and how
the results of the observables, we want to match, evolve with changing
the initial parameters of the simulations (keeping simulation time,
orbit and Galactic potential constant).

The general trend for the evolution of a single simulation with time
can be described as:
\begin{itemize}
\item At each peri-centre passage our object loses mass, through the
  Lagrangian points L1 and L2.
\item Due to the mass-loss, the object expands (formation of tails)
  and the (central) surface brightness diminishes with time.
\item The lost particles will align with the orbit, while moving
  out towards the apo-centre of the orbit, forming tidal tails along
  the orbit.
\item Due to the formation of tidal tails the object will look more
  and more elongated.
\item Particles in the tails have slightly different energies and
  angular momentum than the remaining bound object, i.e., they will
  spread out with time, forming longer and longer tidal tails.
\item as velocities are lower closer to apo-galacticon, the tails will
  contract -- the opposite happens close to peri-galacticon.
\item If sufficient mass is lost \citep[usually more than 90\%; see,
  e.g.,][]{smi13} the object is at the brink of destruction and a
  final peri-centre passage will completely dissolve the object,
  turning it into a pure stellar stream. 
\end{itemize}

Now, for this study, we are more interested in the general trends, how
the initial parameters influence the results of the simulations:
\begin{itemize}
\item The lower the initial mass is, the lower is the final mass of
  the object.  The larger the initial Plummer radius is, the easier it
  is for the object to lose mass and the final mass shrinks with
  increasing initial Plummer radius.
\item The stronger the mass-loss, the more the central surface
  brightness is diminished.
\item In contrast the effective radius of our final object will
  increase, because the nearby tails will be confused with parts of
  the bound object.
\item The velocity dispersion will become smaller with increasing
  mass-loss until sufficient mass is lost, that the unbound particles
  dominate the dispersion measurement, boosting the velocity
  dispersion to higher values with higher mass-loss.
\end{itemize}

We can categorise our results of the simulations in three different
regimes:
\begin{itemize}
\item First, we have the 'bound' regime:  High initial masses or very
  concentrated initial objects (small Plummer radius), will be less
  affected by the Galactic tides.  Sufficient mass is still found as a
  bound object and the observables, we measure, are dominated by this
  bound object.  I.e., the mass is mainly the bound mass, the central
  surface brightness is the one of the remaining bound object, the
  effective radius gets smaller, and we measure a velocity dispersion,
  which is only slightly affected by unbound stars.
\item If the mass-loss is very high, and the object is at the brink of
  total destruction, then the observables are fast changing functions
  of the initial parameters: final mass is dropping fast with larger
  initial radii and/or smaller initial masses, the central surface
  brightness is also dropping fast, the effective radius is increasing
  rapidly, and so does the velocity dispersion, as it is extremely
  boosted by unbound stars.  We dub this the 'intermediate' regime.
  In this regime we expect to find the correct solution. 
\item Finally, if all mass is lost and the object is completely
  destroyed, we are left with a stellar stream without an object.
  This happens at very low initial masses and/or very extended initial
  objects.  We call this the 'stream' regime.  Observables, measured
  in this regime, are almost constant and very low.
\end{itemize}

\subsection{Final Mass}
\label{sec:mass}

Of course, we cannot use the remaining bound mass of our model, as we
see Segue~1 in the final stages of dissolution in our scenario.
Therefore, the total luminosity determined by the observers includes
the faint luminosities of the tails around the object as well.  We, 
therefore, count all particles in a box which spans $\pm 0.20$~degree
in right ascension and $\pm 0.17$~degree in declination from the
centre, which is equivalent to the outermost luminosity contours drawn
by \citet{sim11}. 

In Fig.~\ref{fig:mass} (top left panel), we show the results for the
final mass ($M_{\rm fin}$) of our object.  In the left half of the
panel we plot these results as function of the initial Plummer mass
($M_{\rm pl}$; open symbols).  Simulations with different initial
radii are represented with different symbols (and colours).  Red
triangles represent simulations with initial Plummer radius of
$5.28$~pc, blue squares represent $5.33$~pc, and green circles
$5.41$~pc.  On the right half of the panel we show the final mass
($M_{\rm fin}$) of our object, now as a function of the initial
Plummer radius (filled symbols).  On this side different symbols (and
colours) represent simulations with different initial Plummer masses.
Red triangles represent simulations starting with $M_{\rm pl} = 5
\times 10^{3}$~M$_{\odot}$, blue squares $7 \times
10^{3}$~M$_{\odot}$, and green circles $10^{4}$~M$_{\odot}$. 

In these double-logarithmic plots most if not all results for a given
$R_{\rm pl}$ on the left and for a given $M_{\rm pl}$ on the right can
be fitted with a single straight line, i.e., a power-law function.
These fitting lines are shown in the same colour as their
data-points.  We see that in the right half of the panel the
results in the upper part are deviating from the power-law.  This
behaviour, described already in \citet{bla15}, stems from results
which belong to the first regime of resulting objects, i.e., here we
have bound objects which lose only part of their initial mass.
Furthermore, it is clear that the power-laws on this side of the panel
can not extend to final masses, which are higher than the initial
ones.  Naturally, we exclude those deviating results from the fit.  

Having established that for each constant initial radius and for each
constant initial mass we can fit a power-law to the results (final
mass as function of the initial mass on the left and final mass as
function of the initial scale-length on the right), we can determine
the intersections of these power-laws with the horizontal line
denoting the correct final mass, we assume for Segue~1
($680$~M$_{\odot}$).  This gives us pairs of initial conditions, where
our simulations would lead to the correct final mass for Segue~1.  In
the plot these are denoted by small black squares.  If we now plot all
these possible solutions for the correct final mass into a plot of
initial mass versus initial scale-length as done in the top left panel
of Fig.~\ref{fig:power}, we again note that all these possible
solutions are following a straight line in a double-logarithmic plot,
i.e., they can be fitted by a power-law of the form: 
\begin{eqnarray}
  \label{eq:mass}
  M_{\rm pl} & = & 67.6^{+8.3}_{-7.3} \times R_{\rm
    pl}^{2.60\pm 0.06}. 
\end{eqnarray}
This equation describes the 1D subset of initial parameters (from the
2D parameter space of initial conditions) that leads to the assumed
observed value of the mass of Segue 1.

\subsection{Central surface brightness}
\label{sec:surf}

As the final objects in our simulations span the range from perfectly
bound objects to completely destroyed ones, i.e., pure stellar
streams, there is no simple radial profile, which could fit all the
data of all simulations.  We therefore produce pixel-maps with a
$20$~pc resolution per pixel and determine the brightness of the
densest pixel, which we consider the centre of our object.  As
explained above, objects become larger with simulation time, this
$20$~pc resolution has nothing to do with the initial scale-lengths
used for our models (in the order of $5$ to $6$~pc).  It rather
represents the resolutions found with star count contours of dSph
galaxies observations in the MW \citep[see for example]{irw95}. 

Plotting the central surface brightness of our objects as
function of the initial parameters in Fig.~\ref{fig:mass} (top right
panel, symbols, colours and lines are the same as explained above), we
fit power-laws to the data-points in the intermediate region and
determine again which initial Plummer radius leads to the correct
central surface brightness for a given initial mass and which initial
mass leads to the correct central surface brightness for a given
initial radius.  The data-points obtained from these fits are shown in
the top right panel of Fig.~\ref{fig:power}.  The fitting line
(power-law) to these data points is  
\begin{eqnarray}
  \label{eq:surf}
  M_{\rm pl} & = & 46.8^{+10.7}_{-8.8} \times R_{\rm
    pl}^{2.83 \pm 0.11}. 
\end{eqnarray}
This power-law describes all pairs of initial parameters which
lead to final objects, matching the central surface brightness of
Segue~1. 

\subsection{Effective radius}
\label{sec:rad}

To determine the effective radius we have no other choice 
than to fit single Sersic profiles to our simulation data, without
making a distinction between bound and unbound particles (see
previous section).  We exclude the central region, which may be
dominated by a bound remnant, and we also exclude the far away tidal
tails, as they are too faint to be detected observationally.

Our fits show Sersic indices which are close to $1$.  This is
equivalent to an exponential profile which is observed with most dSph
galaxies.  

We see in Fig.~\ref{fig:mass} (bottom-left panel) that the
final effective radius increases with the initial Plummer radius
(right half).  In the 'bound regime' the increase is slow and
reflects the simple fact that we start with more extended objects.  In
the intermediate region the effective radius increases fast as more
and more material is in the tidal tails and the remaining bound object
expands as well.  Finally, once only a stream is left, the increase
levels off as now we measure the extent of the stream rather than a
meaningful profile.

Using the same procedure as described before we show the resulting
data in the lower left panel of Fig.~\ref{fig:power} and describe the
fitting line as:
\begin{eqnarray}
  \label{eq:rad}
  M_{\rm pl} & = & 39.81^{+6.96}_{-5.93}\times R_{\rm
    pl}^{2.88 \pm 0.10}. 
\end{eqnarray}

\subsection{Velocity dispersion}
\label{sec:vel}

To determine the velocity dispersion, we calculate the total velocity
dispersion using the radial velocities of all particles within the
same box, we used to determine the final mass.  This is also the
region, where most of the stars are found, which were used to
determine the velocity dispersion observationally. 

The results in Fig.~\ref{fig:mass} (bottom-right panels) show for a
given initial mass as function of the initial radius first a
decreasing dispersion with increasing radius and then a turn around
and an increase steeply with increasing initial radius (right-panel).
This behaviour can be understood as follows:  First, we measure the
velocity dispersion of the bound object (with only a small
contamination of unbound stars), which decreases as the final object
is less and less massive.  At one point the velocity dispersion gets
dominated by unbound stars and is increasing again, as unbound stars
are on different orbits around the Galaxy and we measure a
distribution of epicyclic frequencies rather than a ’real’ velocity
dispersion.  In this region, where the velocity dispersion is boosted
by unbound particles, we determine our power-laws.  The results 
are shown in the lower right panel of Fig.~\ref{fig:power} and can be 
described as:
\begin{eqnarray}
  \label{eq:vel}
  M_{\rm pl} & = & 50.1^{+16.0}_{-12.1} \times R_{\rm
    pl}^{2.77 \pm 0.16}. 
\end{eqnarray}

One could now ask how robust is our measurement of the velocity
dispersion and if we do include particles, which observers would
reject as interlopers or unbound, thereby inflating our dispersion to
the desired high values.  We repeat the analysis on our best fitting
model (see next section) with different observational methods as
described in \citet{smi13} and find all methods to lead to extremely
boosted velocity dispersions.

It is very easy for us to determine which particles of our simulation
are still bound to each other.  If we determine their line-of-sight
velocity dispersion of the bound particles alone (and only with this
velocity dispersion the virial theorem applies) we obtain $\sigma_{\rm
  bound} = 0.29$~km\,s$^{-1}$.  This shows clearly that velocity
dispersions can be severely boosted by unbound particles.  

With the method we use, i.e., to take the projected velocities of
all the particles in the same box where we measure the final mass,
our analysis gives us a velocity dispersion of $\sigma_{\rm los} =
3.52$~km\,s$^{-1}$.  Now, our sample only contains particles ('stars')
which belong or used to belong to the object.  In a simulation we do
not have contamination from random halo stars, which happen to be in
the same field of view.  We mimic a three-sigma-clipping algorithm
used by \citet{YV77} to reject interlopers (which are not present) and
obtain a somewhat lower dispersion of $\sigma_{\rm clip} =
2.20$~km\,s$^{-1}$.  If we apply the new interloper rejection
technique (IRT) described by \citet{Klimen07}, we arrive at a
somewhat higher value of $\sigma_{\rm IRT} = 3.91$~km\,s$^{-1}$.  Even
though those three methods do not agree very well with each other,
they all show a boosted dispersion by an order of magnitude compared
to the bound particles alone.  As shown in e.g.,\ \citet{smi13} the
boosting is higher the closer the object is to destruction and the
closer we see the object to its apo-galacticon, when the tails get
compressed.  Both effects are at work at our solution for Segue~1.  

\subsection{Best fitting model}
\label{sec:best}

\begin{figure*}
  \centering
  \includegraphics[width=8cm, height=8cm]{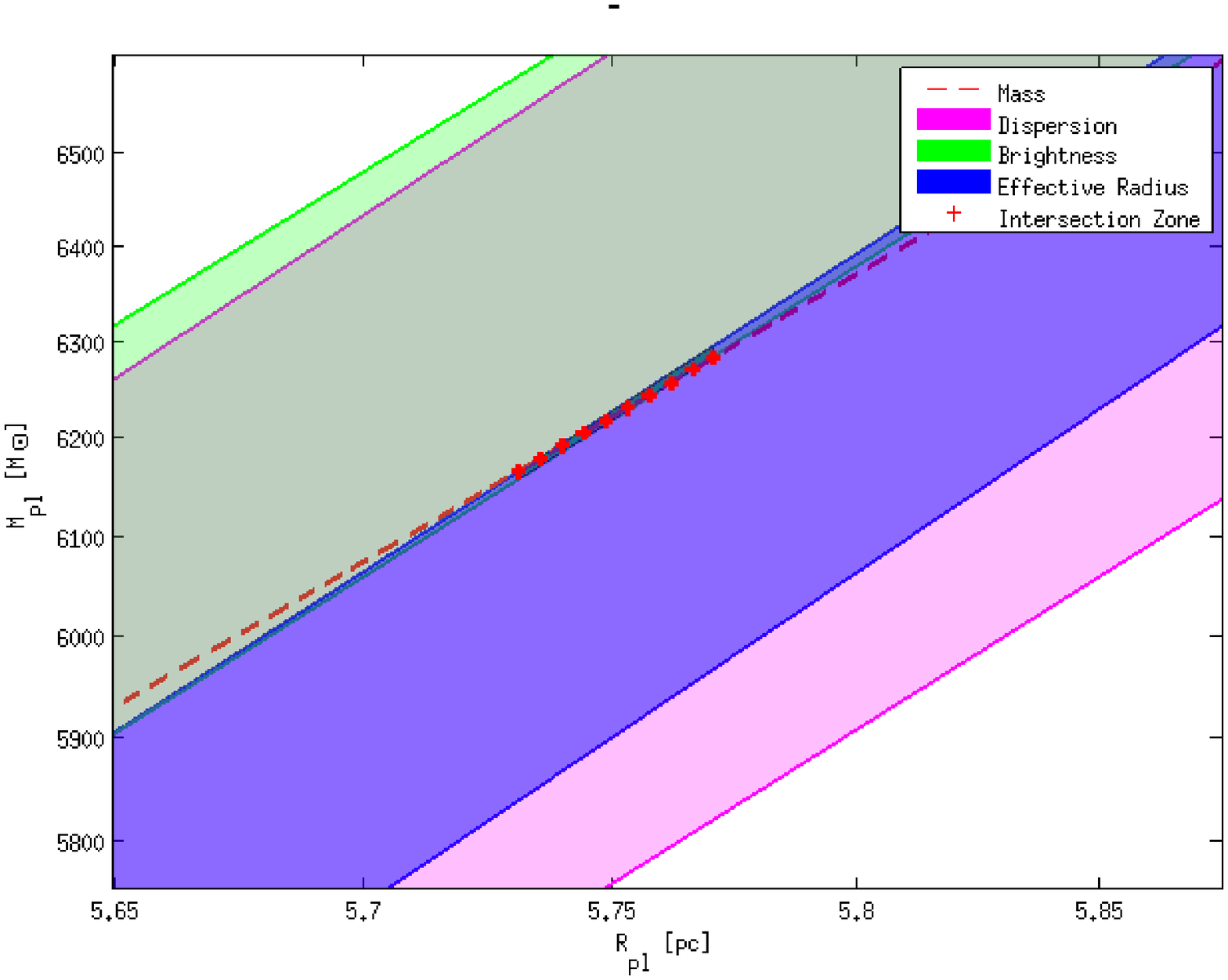}
    \includegraphics[width=8cm, height=8cm]{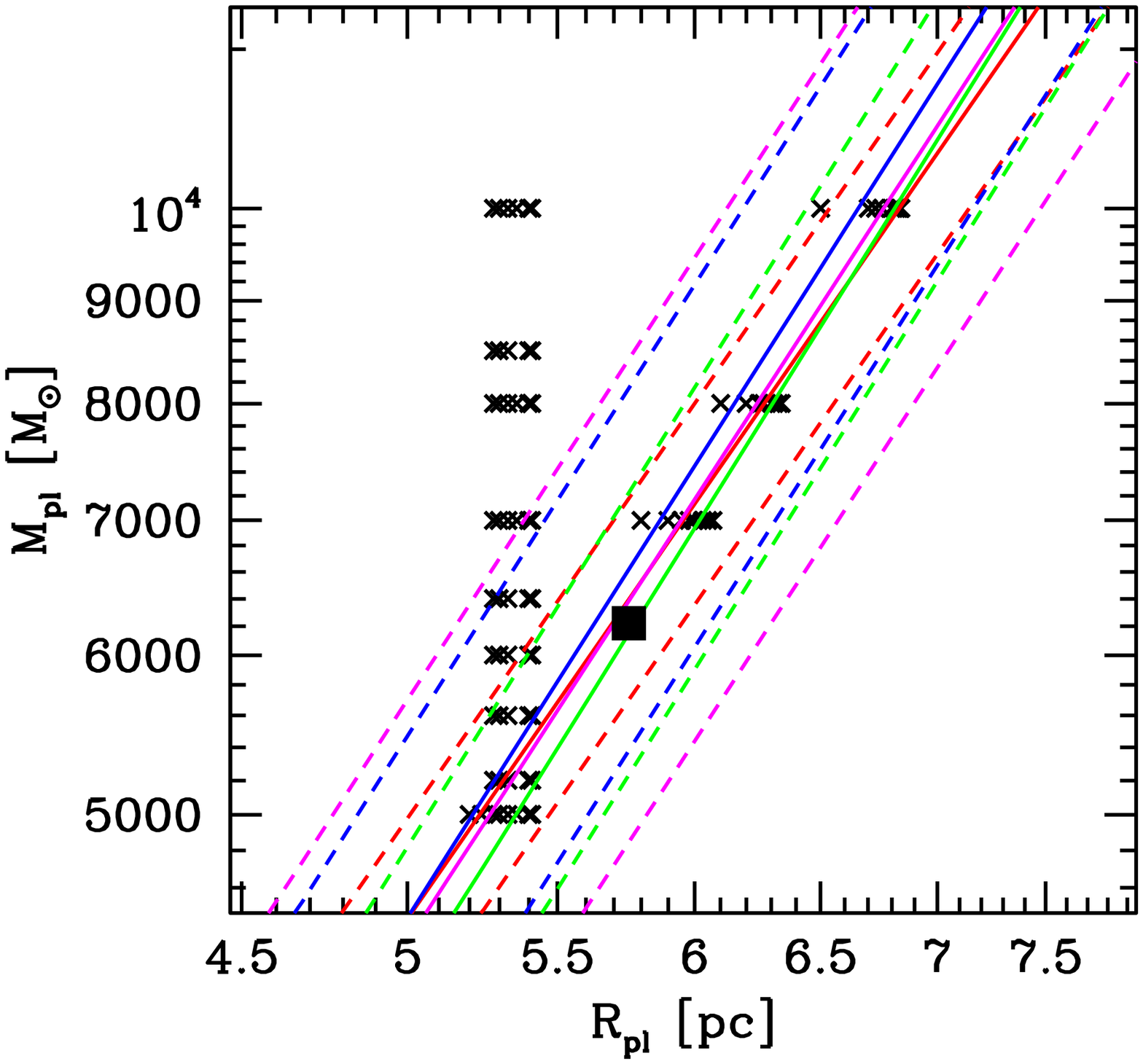}
  \caption{Joining all power-law fits from Fig.~\ref{fig:power} in one
    single graph.  Left panel shows shaded areas representing the
    observational uncertainties for the observational quantities
    (except mass here we cannot address the error; see main text).  In
    the right panel we show the fitting lines (solid) together with
    the mathematical errors ($1\sigma$) as dashed lines.  Red lines for
    the mass, blue for the surface-brightness, green for the effective
    radius and magenta for the velocity dispersion.  In an ideal world
    all lines should intersect in one point, where the correct initial
    conditions of Segue~1 should be found.  Instead the lines intersect
    inside a small region.  In the left panel this region is denoted
    with red crosses; in the right panel we show a big black square
    where we place our best fitting model.  Furthermore, the right
    panel shows all initial conditions of our simulations as black
    crosses.}  
  \label{fig:fit}  
\end{figure*}

\begin{figure}
  \centering
\includegraphics[width=9cm, height=9cm]{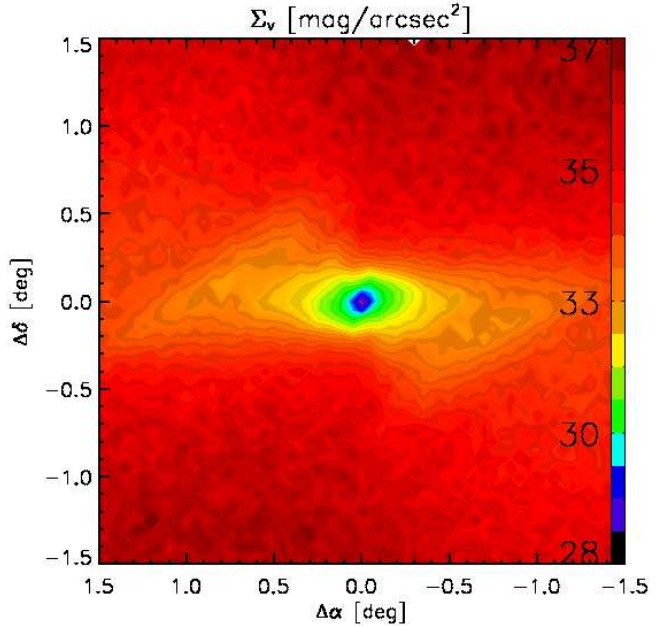}  
  \caption{Surface brightness contours of our best fitting model,
    resembling Segue~1.} 
  \label{fig:final}
\end{figure}

Having established how the four observables depend on our initial
parameters, we now have a look, if our solutions converge towards a
single solution.  We show the different fitting lines from
Fig.~\ref{fig:power} together in the panels of Fig.~\ref{fig:fit}.
We see clearly that the lines intersect but not at the same point.
This would only be the case in an absolute ideal world, where none of
the observables would have any error margin (and of course if a
solution exits).  Taking the one sigma deviations into account,
published for the different observables (excluding final mass as here
we have no handle on the error stemming from determining the total
visual magnitude and then transforming it into a mass using a rather
generic mass-to-light ratio), we obtain a small region in initial
parameter space in which all observational quantities are matched
within their observational one sigma errors (shown as red crosses in
the right panel of Fig.~\ref{fig:fit}).  We place our best-fitting
model in the centre of this small area using an initial Plummer radius
of $R_{\rm pl} = 5.75$~pc and an initial mass of $M_{\rm pl} =
6224$~M$_{\odot}$.  

Running a last simulation with these initial parameters leads to a
final object for which we show the surface brightness contours in
Fig.~\ref{fig:final}.  It has the correct elongation and shows faint
tails, which may have been picked up in the study of \citet{nie09}.
The blue regions in the figure represent surface brightnesses,
brighter than $30$~mag\,arcsec$^{-2}$, which are now possible to
observe via star counts.  The green, yellow and orange regions have
very low surface brightnesses and may get picked up by chance.  They
are located in the same places as the tentative patches of
\citet{nie09}.  All the red and dark red regions have brightnesses
fainter than about $35$~mag\,arcsec$^{-2}$ and are impossible to
observe.  In our simulations we have maybe a handful of particles
(phase-space elements) per pixel in those areas, which would represent
less than one observable star per pixel in reality.  

We obtain for this best fitting model the following 'observables':  
\begin{enumerate}
\item A final mass of $686^{+28}_{-52}$~M$_{\odot}$ measured in the
  region described above.
\item A central surface brightness of $\mu_{0} =
  28.4^{+0.8}_{-0.3}$~mag. 
\item An effective radius of $26.6^{+3.8}_{-3.7}$~pc.
\item A line-of-sight velocity dispersion of
  $3.29^{+0.12}_{-0.31}$~km\,s$^{-1}$. 
\end{enumerate}
These values are in very good agreement with the observed quantities
shown in Table~\ref{tab:obs}.

\section{Conclusion \& Discussion}
\label{sec:conc}

We have embarked on a project to find a non dark matter dominated
progenitor for the Segue~1 dSph.  With our simulations we could show
that assuming an orbit with proper motions of $\mu_{\alpha} =
-0.19$~mas\,yr$^{-1}$ and $\mu_{\delta} = -1.9$~mas\,yr$^{-1}$
($R_{\rm peri} = 2.9$~kpc, $R_{\rm apo} = 31.7$~kpc), which places the
object near its apo-galacticon today, we are indeed able to reproduce
all observables dealing with kinematics.    

The choice of our orbit relies heavily on our assumption that the
patches of stars observed by \citet{nie09} are indeed parts of the
tidal tails of Segue~1.  If this assumption is false, then the choice
of orbit becomes completely arbitrary (as long as the orbit matches
the determined radial velocity).  On the other hand, assuming
any orbit would make it much easier to find a suiteable one in which
we see Segue~1 close to its apo-galacticon, at the final stages of
destruction and on top of that in the favourable position that we look
partly along the tidal tails to have a better boosting of the velocity
dispersion.  So, while our restriction of the path of the orbit leads
to a precise prediction of the initial conditions, we need to
reproduce the observables of Segue~1, a total freedom of the orbit
would give us more and easier possibilities to find matching
solutions.  

Using a Plummer sphere as initial model, we find that the progenitor
of Segue~1 can be a small star cluster that formed $10$~Gyr ago with a
scale-length of less than $6$~pc and an initial mass of about $6
\times 10^{3}$~M$_{\odot}$.  Today, Segue~1 has, according to our model,
lost most of its mass, it is at the brink of destruction, and it is
located close to its apo-galacticon.  Therefore, we see an extremely
boosted velocity dispersion (more than an order of magnitude) stemming
almost entirely from unbound stars. 

This study shows that from a kinematical and structural point of view,
it is indeed possible to explain Segue~1 as a completely dark matter
free entity, without invoking new physics like MOND.  It is, on the
other hand, very straightforward to explain Segue~1 as a highly DM
dominated object.  One just takes the kinematical observables, assumes
virial equilibrium and obtains a mass-to-light ratio which is
impossible to explain by a pure stellar population alone.  Models like
\citet{ass13a,ass13b} show that most if not all kinematical
peculiarities of dSph galaxies (e.g., elongations, off-centre nuclei,
twisted contours, cold sub-populations, etc.) can be explained by
clustered star formation inside of DM haloes.  These models work
without the need for any additional perturbations from the MW or other
galaxies.  It is therefore not difficult to explain Segue~1 as a DM
dominated dwarf galaxy, and even though this might be the correct
answer, this in not part of this study, which is looking for the
possibility to find an alternative explanation, which we have.

Our model lacks one big ingredient.  We are using a particle-mesh code
to simulate a small star cluster, thereby neglecting completely the
internal evolution of the star cluster and the resulting mass-loss due
to two-body relaxation.  These fast models allow us to cover a vast
parameter space of initial conditions on various orbits, while direct
N-body simulations, even though possible for objects similar to our
initial models, are time-consuming and require special hardware.

\citet{fel07} showed that, in the case of NGC~5466, the mass-loss due
to the Galactic potential is about 60\% of the the total mass lost,
including two-body relaxation and stellar evolution.  In the case of
Segue~1, due to its orbit closer to the Galactic centre, we expect
this percentage to be even higher.  Nevertheless, one should regard
our initial mass of Segue~1 as a lower limit, as it neglects stellar
evolution and relaxation effects.  The 'birth' mass may even be much
higher, as stellar evolution in the first few Myr is fast and coupled
with a strong mass-loss (e.g.,\ Supernovae and/or stellar winds of high
mass stars).  If additionally effects like gas-expulsion of an
embedded young cluster is taken into account, depending on a
completely unknown star formation efficiency, the real initial mass is
completely unknown.  Our initial mass represents, at best, a lower
limit for a gas-free star cluster, after the violent and fast initial
evolution, i.e., a couple of tens of Myr after its birth.

One other caveat, is the metalicity spread observed in Segue~1, which
is higher than expected from any DM free star cluster.  In our study
we focus on the kinematics alone.  In a particle-mesh code, particles
represent phase-space elements and not stars.  Therefore, this study
cannot give an explanation for any star formation and/or chemical
enrichment history of Segue~1.  Our study only shows, that it is
possible to reproduce all kinematical and structural properties of
Segue~1 with a DM free model. 
\\

{\bf Acknowledgments:}
MF acknowledges financial support of FONDECYT grant no.\ 1130521 and
BASAL PFB-06/2007. RD and JPF are partly funded through a studentship
of FONDECYT grant no.\ 1130521.  JPF acknowledges funding through a
CONICYT studentship for Magister students.  JD is funded through
FONDECYT grant no. 3140146 and GNC is funded through FONDECYT grant
no. 3130480. 

\bibliographystyle{mnras}

\begin{thebibliography}{xxx99}

\bibitem[\protect\citeauthoryear{Assmann et al.}{2013a}]{ass13a}
  Assmann P., Fellhauer M., Wilkinson M.I., Smith R. 2013, MNRAS, 432,
  274 

\bibitem[\protect\citeauthoryear{Assmann et al.}{2013b}]{ass13b}
  Assmann P., Fellhauer M., Wilkinson M.I., Smith R., Bla\~{n}a
  M. 2013, MNRAS, 435, 2391 

\bibitem[\protect\citeauthoryear{Baltz et al.}{2000}]{bal00}
  Baltz E.A., Briot C., Salati P., Taillet R., Silk J. 2000, PhysRevD,
  61, 3514

\bibitem[\protect\citeauthoryear{Belokurov et al.}{2007}]{bel07}
  Belokurov  V., et al. 2007, ApJ, 654, 897

\bibitem[\protect\citeauthoryear{Binney \& Tremaine}{1987}]{bin87}
  Binney J., Tremaine S. 1987, 'Galactic Dynamics', Princeton Series
  in Astrophysics, ISBN 0-691-08445-9

\bibitem[\protect\citeauthoryear{Bla\~{n}a et al.}{2015}]{bla15}
  Bla\~{n}a M., Fellhauer M., Smith R., Candlish G.N., Cohen R.,
  Farias J.P. 2015, MNRAS, 446, 144

\bibitem[\protect\citeauthoryear{Boily \& Kroupa}{2003}]{boi03}
  Boily C.M., Kroupa P. 2003, MNRAS, 338, 673

\bibitem[\protect\citeauthoryear{Evans, Ferrer \& Sarkar}{Evans et
    al.}{2004}]{eva04}
  Evans N.W., Ferrer F., Sarkar S. 2004, PhysRevD, 69, 123501

\bibitem[\protect\citeauthoryear{Fellhauer et al.}{2000}]{fel00}
  Fellhauer M., Kroupa P., Baumgardt H., Bien R., Boily C.M., Spurzem
  R., Wassmer N. 2000, New Ast., 5, 305

\bibitem[\protect\citeauthoryear{Fellhauer et al.}{2007}]{fel07}
Fellhauer M., Evans N.W., Belokurov V., Wilkinson M.I., Gilmore
G. 2007, MNRAS, 380, 749

\bibitem[\protect\citeauthoryear{Frebel, Simon \& Kirby}{Frebel et
    al.}{2014}]{fre14} 
  Frebel A., Simon J.D., Kirby E.N. 2014, ApJ, 786, 74

\bibitem[\protect\citeauthoryear{Geha et al.}{2009}]{geh09}
  Geha M., Wilman B., Simon J.D., Strigari L.E., Kirby E.N., Law D.R.,
  Strader J. 2009, ApJ, 692, 1464 

\bibitem[\protect\citeauthoryear{Irwin \&
    Hatzidimitriou}{1995}]{irw95}
  Irwin M., Hatzidimitriou D. 1995, MNRAS, 277, 1354

\bibitem[\protect\citeauthoryear{Kuhlen et al.}{2008}]{kuh08}
  Kuhlen M., Diemand J., Madau P., Zemp M. 2008, Journal of Physics:
  Conference Series, Vol.\ 125, Issue 1, p.1

\bibitem[\protect\citeauthoryear{Laevens et al.}{2015}]{lae15}
  Laevens B.P.M. et al. 2015, ApJ submitted, arXiv:1507.07564

\bibitem[\protect\citeauthoryear{Martinez et al.}{2009}]{mar09}
  Martinez G.D., Bullock J.S., Kaplinghat M., Strigari L.E., Trotta
  R. 2009, JCAP, 06, 14

\bibitem[\protect\citeauthoryear{Martin et al}{2007}]{mar07}
  Martin N.F., Ibata R.A., Chapman S.C., Irwin M., Lewis G.F. 2007,
  MNRAS, 380, 281

\bibitem[\protect\citeauthoryear{Martin, de Jong \& Rix}{Martin et
    al.}{2008}]{mar08} 
  Martin N.F., de Jong J.T.A., Rix H.-W. 2008, ApJ, 684, 1075

\bibitem[\protect\citeauthoryear{Mizutani, Chiba \& Sakamoto}{Mizutani
    et al.}{2003}]{miz03}
  Mizutani A., Chiba M., Sakamoto T. 2003, ApJ, 589, 89

\bibitem[\protect\citeauthoryear{Niederste-Ostholt et
    al.}{2009}]{nie09}
  Niederste-Ostholt M., Belokurov V., Evans N.W., Gilmore G., Wyse
  R.F.G., Norris J.E. 2009, MNRAS, 398, 1771

\bibitem[\protect\citeauthoryear{Pieri L. et al.}{2009}]{pie09}
  Pieri L., Pizella A., Corsini E.M., Dalla Bont\`{a} E., Bertola
  F. 2009, A\&A, 496, 351

\bibitem[\protect\citeauthoryear{Plummer}{1911}]{plu11}
  Plummer H.C. 1911, MNRAS, 71, 460

\bibitem[\protect\citeauthoryear{Simon et al.}{2011}]{sim11}
  Simon J.D. et al. 2011, ApJ, 733, 46

\bibitem[\protect\citeauthoryear{Smith et al.}{2013}]{smi13}
  Smith R., Fellhauer M., Candlish G.N., Wojtak R., Farias J.P.,
  Bla\~{n}a M. 2013, MNRAS 398, 1771

\bibitem[\protect\citeauthoryear{York et al.}{2000}]{yor00}
  York D.G., et al. 2000, AJ, 120, 1579

\bibitem[\protect\citeauthoryear{Klimentowski et al.}{2007}]{Klimen07} 
  Klimentowski, J., et al. 2007, MNRAS, 378, 353 

\bibitem[\protect\citeauthoryear{Yahil \& Vidal}{1977}]{YV77} 
Yahil, A., \& Vidal, N.~V. 1977, ApJ, 214, 347 

\end{thebibliography}

\label{lastpage}
\end{document}